\documentclass[smallextended]{svjour3}
\usepackage{graphicx}
\usepackage{url}

\newcommand{\urms}[1]{u_{\textnormal{rms}}}

\makeatother

\begin{document}
\title{Background Oriented Schlieren in a Density Stratified Fluid}
\author{Lilly Verso \and Alex Liberzon}

 \institute{ Turbulence Structure Laboratory, School of Mechanical Engineering,\\
Tel Aviv University, Tel Aviv 69978, Israel}

\date{}

\maketitle

\begin{abstract}
Non-intrusive quantitative fluid density measurements methods are essential in stratified flow experiments. Digital imaging leads to synthetic Schlieren methods in which the variations of the index of refraction are reconstructed computationally. In this study, an important extension to one of these methods, called Background Oriented Schlieren (BOS), is proposed. The extension enables an accurate reconstruction of the density field in stratified liquid experiments. Typically, the experiments are performed by the light source, background pattern, and the camera positioned on the opposite sides of a transparent vessel. The multi-media imaging through air-glass-water-glass-air leads to an additional aberration that destroys the reconstruction. A two-step calibration and image remapping transform are the key components that correct the images through the stratified media and provide non-intrusive full-field density measurements of transparent liquids.     
\end{abstract}

\section{Introduction}\label{sec:introduction}

Non-invasive measurements of density are of special importance for stratified fluids experiments. Over the last decades numerous optical techniques were developed \cite{tropea_yarin_foss:2007}. For instance, Schlieren, Shadowgraphy, Interferometry, and others, allow for density variation measurements based on the differences in the fluid refractive index \cite{Settles}. Progress in digital imaging helped to create the digital counterparts, such as synthetic Schlieren methods \cite{Dalziel2000,Richard1998}.

The refractive index varies as $n=c/c_{0}$, where $c, c_{0}$ are 
the light velocity in the medium and in the vacuum, respectively. Light ray
passing through a non-uniform density region is deflected from its original
path. The relation between the density inhomogeneity and the refractive index
are defined by Gladstone-Dale  $n-1 = G(\lambda)\, \rho$, where $G(\lambda)$ is the Gladstone-Dale constant, $\lambda$ is the wavelength of a light beam and $\rho$ is the density of the fluid \cite{Goldstein}.

The Background Oriented Schlieren belongs to the family of synthetic Schlieren methods \cite{Dalziel2000,Richard1998,Venkatakrishnan2004}. The BOS exploits a light source that projects a textured background (typically a random pattern of dots) located on one side of a test chamber onto the camera sensor positioned on the opposite side. The first image (called reference image) is recorded on the background pattern through a stagnant fluid of uniform density. Upon fluid motion or density variations, density gradients induce the distortions of the projected pattern compared to the reference image. The distortions are quantified using optical flow or particle image velocimetry (PIV) methods, and provide the field of virtual displacements proportional to the derivatives of the index of refraction \cite{Dalziel2000,Richard1998}.

Dalziel and co-authors~\cite{Dalziel2000} utilized the synthetic Schlieren methods to calculate the gradient of the density fluctuations in an air flow around a cylinder. Generally, the majority of the BOS applications are in aerodynamic applications. Raffel \& Richard~\cite{Richard1998} performed an intensive BOS investigation on the formation and interaction of the blade vortex phenomenon, aiming to reduce the noise generated by a rotor of a helicopter. Validation of the BOS technique can be found in Venkatakrishnan \& Meier~\cite{Venkatakrishnan2004}. The authors compared the 2D-density field of a cone traveling in air at Mach 2 with the known cone-charts. The authors also integrated the field of displacements applying the Poisson equation and reconstructed the density field. Recently, BOS was extended to three-dimensional measurements for air flows by Berger et al.~\cite{Berger}. The authors extended the image processing algorithms to capture time resolved unsteady gas density fields. Over the last decade several studies have focused on the accuracy of the technique. For instance, Elsinga~\cite{Elsinga} and Vinnichenko~\cite{Vinnichenko} tested the influence of various parameters on the measurement sensitivity and resolution of the technique, and suggested a set of guides for an optimal set-up of a BOS system.

To the best of our knowledge, the BOS measurements in stratified liquid flows are limited to the density gradients only. For instance, Sutherland~\cite{Sutherland:1999} tracked internal waves using the fields of density gradients. The reconstruction of the density field, based on the Poisson equation to the density field, is not found in the literature. 

In this study we note that the distortions arise from the several sources in the multimedia (air-glass-liquid-glass-air) imaging, specifically in its digital version. The air-glass-liquid interface with sharp refractive index changes its behavior as a lens that amplifies the optical aberrations of the light source, camera and lenses. Therefore, it is necessary to calibrate the BOS optical system through the here  proposed multi-step method. The procedure starts with the BOS pattern image through air in the test-section (air-glass-air-glass-air), followed by a homogeneous liquid calibration (air-glass-liquid-glass-air) and digital image remapping. 

The new multi-step calibration procedure is accompanied by a novel digital image remapping method which
corrects the displacements field. The remapping is based on the displacement field, which is obtained by correlation of the reference image and the image of a homogeneous liquid. This step is followed by the reference image captured through a stagnant stratified fluid. The important result of this study is that the calibration and digital image correction allow us to reconstruct the correct density field based on solution of the Poisson equation. We validate the method correctly reconstructing the density of two tests: a) air-water interface; and b) the multi-layer stably stratified saline solution. 

The paper is organized as follows. In  Section~\ref{sec:principles} we briefly review the relevant principles of the background oriented Schlieren method. Section~\ref{sec:algorithms} explains the image processing and reconstruction algorithms applied to the experimental data. Section~\ref{sec:setup} shows the experimental setup and the experimental results of the two tests. Finally, we summarize the study in Section~\ref{sec:conclusions}.

\section{BOS model}\label{sec:principles}

In this section the basic principles of the BOS technique are summarized for the sake of brevity and augmented by an extension that allows reconstructing fluid density in stratified liquids. 

\begin{figure}[ht]
\centering\includegraphics[width=.8\textwidth]{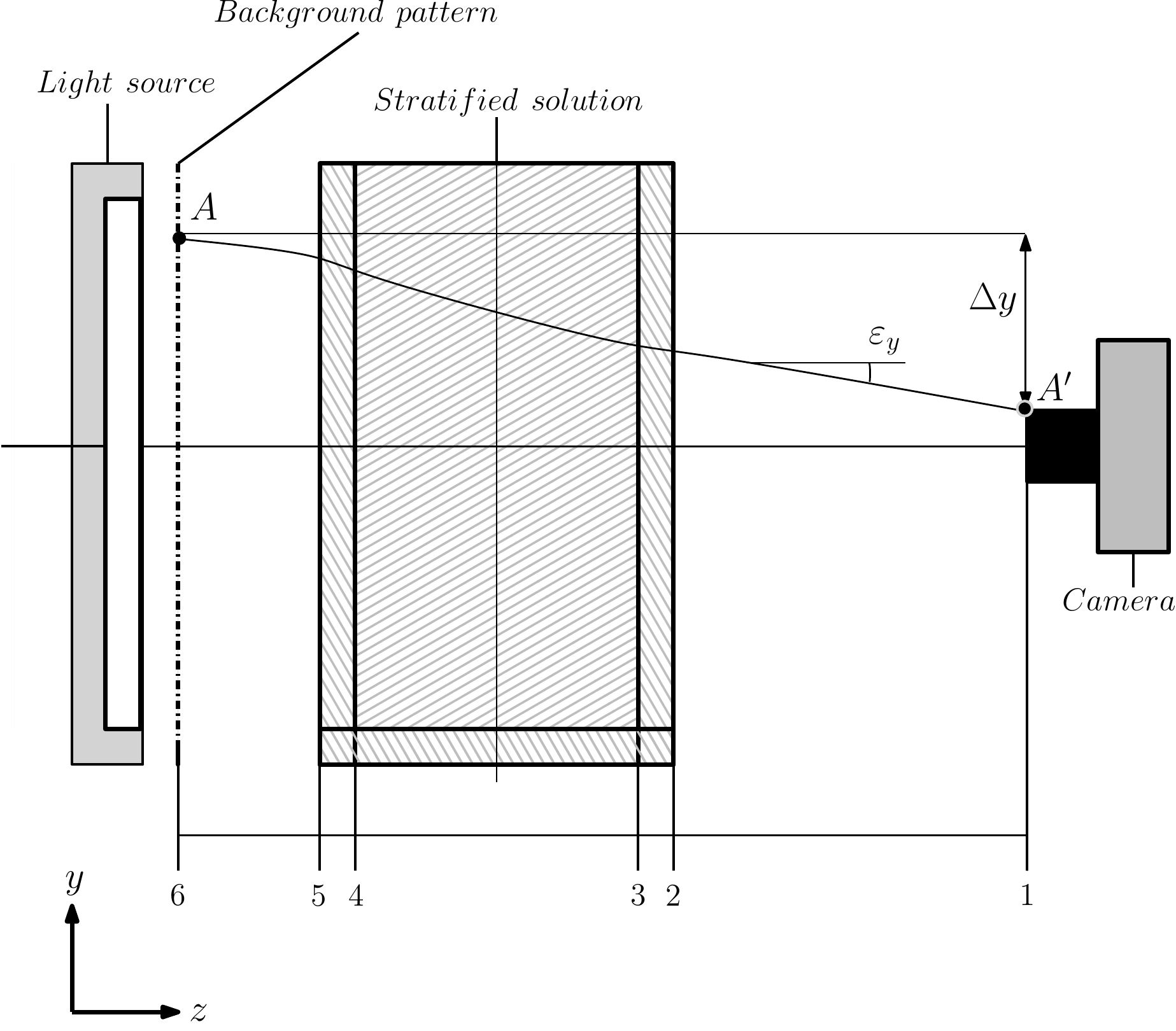}
\protect\caption{Schematic view of a BOS setup. $z$-axis is the imaging axis and $x,y$ are the orthogonal coordinates. Important cross-sections are marked 1-6. \label{fig_system}}
\end{figure}

The common setup is a background random dots pattern illuminated by a light source and a digital camera facing the light passing the fluid (gas or liquid), as shown schematically in Fig.~\ref{fig_system}. 

\begin{figure}[ht]
\centering \includegraphics[width=.8\textwidth]{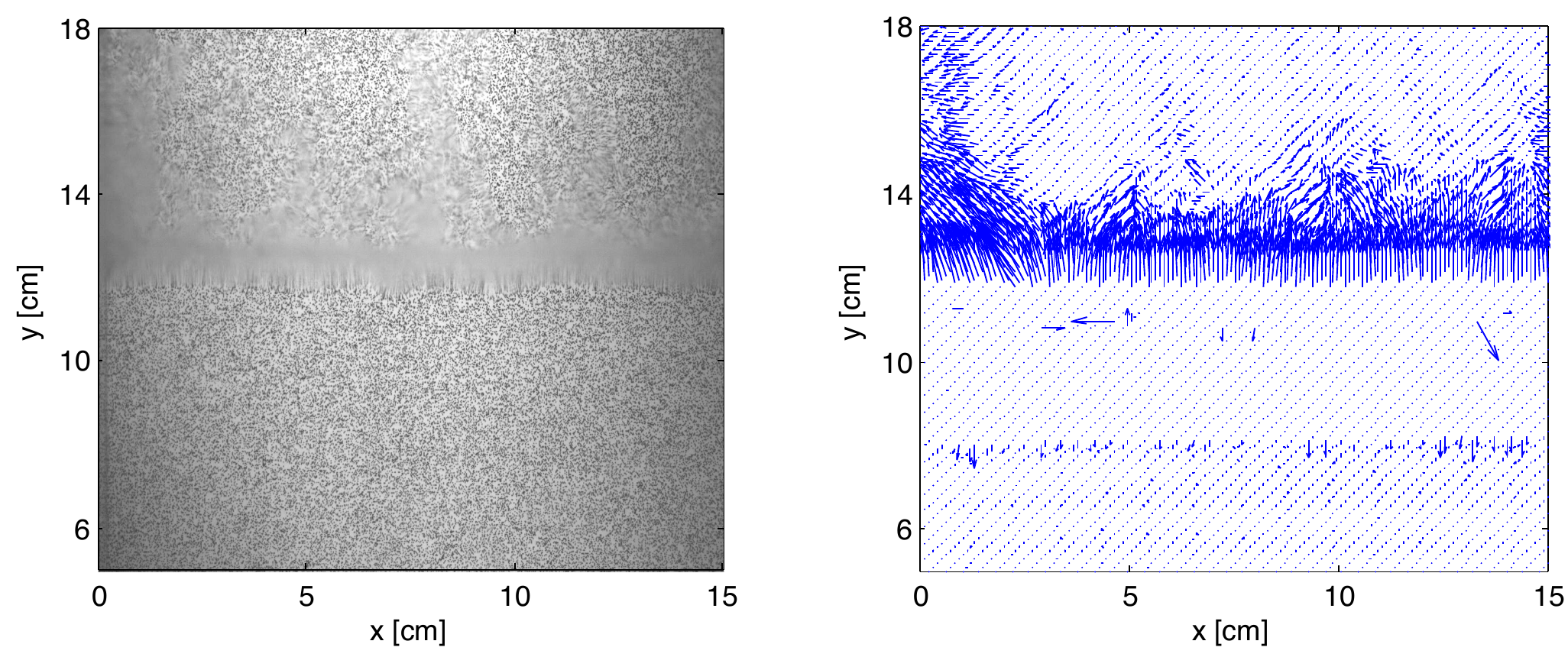}
\caption{(a) Distorted background image; and (b) the corresponding displacement field in a setup of a water layer above a layer of  saline solution. \label{fig_displ}} 
\end{figure}

The BOS technique is based on the distortion of the background image due to density
changes. The image is distorted with respect to the reference image of the random pattern, without the fluid. The distortion is a cumulative effect of the refractive index variation along the light ray passing through the fluid. The displacement fields $\Delta x_{d}$ and $\Delta y_{d}$ can be calculated based on the correlation of the reference and the distorted images. For example, Fig.~\ref{fig_displ} demonstrates the distorted image in our experiment and the corresponding displacement field.   

Light ray crossing the interface of different indices of refraction changes the direction (refracts) proportional to the ratio of the indices, based on Snell's law. Consequently, a curvature of the light ray can be approximated as in Settles \cite{Settles}:
%
%
%
%
\begin{equation} \label{int_1}
\frac{d^{2}x}{dz^{2}}  =  \frac{1}{n}\frac{\partial n}{\partial x}\qquad 
\frac{d^{2}y}{dz^{2}}  =  \frac{1}{n}\frac{\partial n}{\partial y}
\end{equation}
\noindent Integrating  Eq.(\ref{int_1}) for a thickness of a fluid layer, $W$, the deflection angles $\alpha_{x}$ and $\alpha_{y}$ are obtained \cite{Raffel2015}:
\begin{equation}
\tan\alpha_{x}  =  \frac{dx}{dz}=\int\limits_0^{H} \frac{1}{n}\frac{\partial n}{\partial x}\ dz \qquad
\tan\alpha_{y}  =  \frac{dy}{dz}=\int\limits_0^{H} \frac{1}{n}\frac{\partial n}{\partial y}\ dz
\end{equation}

When the BOS is implemented for gas flows or for the density gradient field, these models are sufficient for further analysis\cite{Settles,Raffel2015}. However,  in the specific case of our interest, light rays refract also at the air-glass, glass-liquid, liquid-glass and glass-air interfaces. According to Fig.~\ref{fig_system}, it passes through  sections 1-5 (number of layers  in our case is $N = 6$). Therefore, an extended analysis is required in order to reconstruct the density field. For every layer of air, glass or liquid,  of thickness say $H_i$, the displacement is estimated according to Eq.(\ref{int_1}):
\begin{equation}
\Delta x_{i} = \int_{0}^{H_i}\bigg[\int_{0}^{H_i}\frac{1}{n_{i}}\frac{\partial n_{i}}{\partial x}dz\bigg]dz=H_i^{2}\frac{1}{n_{i}}\frac{\partial n}{\partial x} 
\end{equation}
 \noindent and $\Delta y_{i}$ obtained equivalently. The total displacement $\Delta x, \Delta y$ is the sum of the individual deflections: 
\begin{equation} \label{dispTot_1}
\Delta x = \sum\limits_{i=1}^{N}  H_i^{2}\frac{1}{n_{i}}\frac{\partial n}{\partial x}   \qquad \Delta y = \sum\limits_{i=1}^{N}  H_i^{2}\frac{1}{n_{i}}\frac{\partial n}{\partial y}
\end{equation}

The following step of the analysis is the reconstruction of the density field based on the Poisson equation \cite{Venkatakrishnan2004}:
\begin{equation}\label{eq_Poisson} 
\Delta n=\frac{\partial^{2}n}{\partial x^{2}}+\frac{\partial^{2}n}{\partial y^{2}}= K \left[\frac{\partial}{\partial x}\Delta x +\frac{\partial}{\partial y}\Delta y \right]
\end{equation}
\noindent where the multiplier $K$ is the inverse of the contributions of different layers:
\begin{equation}
K = \left[2 \sum\limits_{i=1}^{N}  \frac{H_i^{2}}{n_{i}} \right]^{-1}
%
\end{equation}

The BOS method reverses the use of the Poisson equation Eq.~\ref{eq_Poisson}. The displacements $\Delta x, \Delta y$ are first obtained for each coordinate $x,y$ using the cross-correlation PIV algorithm.   Then the gradient of the displacement field is numerically estimated (using high order accuracy numerical methods) and the Poisson equation is solved for an unknown $n_i$. In our case the stratified fluid layer index of refraction $n_3$ is the required result.  

\begin{figure}[ht]
\centering\includegraphics[width=.6\textwidth]{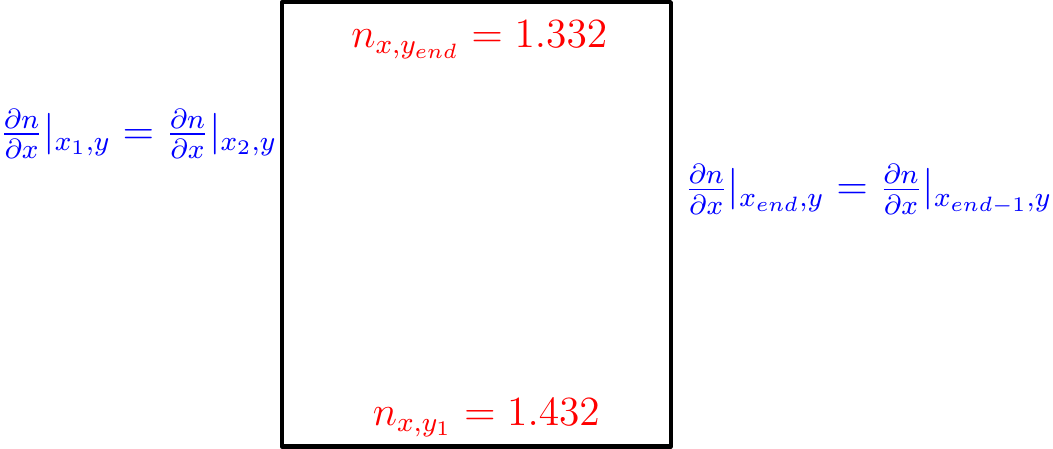}
\caption{Boundary conditions for the Poisson equation solution: top/bottom Dirichlet boundary conditions  $n_{x,y_{1}}=1.432$ and $n_{x,y_{end}} =1.332$. Left/right conditions are Neumann type $\frac{\partial n}{\partial x}|_{x_{1},y}=\frac{\partial n}{\partial x}|_{x_{2},y}$ and  $\frac{\partial n}{\partial x}|_{x_{end},y}=\frac{\partial n}{\partial x}|_{x_{end-1},y}$ \label{fig_bc}} 
\end{figure}

Boundary conditions are necessary for the numerical solution of a partial differential equation. Schematically, the boundary conditions are summarized in Fig.~\ref{fig_bc}. 

Eventually, the Gladstone-Dale relation ($n-1 = G(\lambda)\, \rho$) is used to transform the field of $n$ to the density field $\rho(x,y)$.

\section{Image remapping method}\label{sec:algorithms}

Literature reviews revealed the difficulty to implement the 
synthetic Schlieren method to the stratified fluids. We have identified that the key problem relates to the multi-media imaging path. The method proposed here is called an image remapping method. The method utilizes a multi-step calibration and image processing routine known as remapping. Remapping is the shift of each pixel in the image by a distance prescribed by the displacement field. 

Two reference images are captured when the tank is full with air and water (or another liquid of a uniform index of refraction, close to the final solution).  Liquid in the tank causes an apparent displacement of the dots (referred to the initial image) in the image recorded by the camera. In short, the method explained here separates the result from this apparent distortion using calibration. 
 
\begin{figure}[ht]
\centering\includegraphics[width=0.5\textwidth]{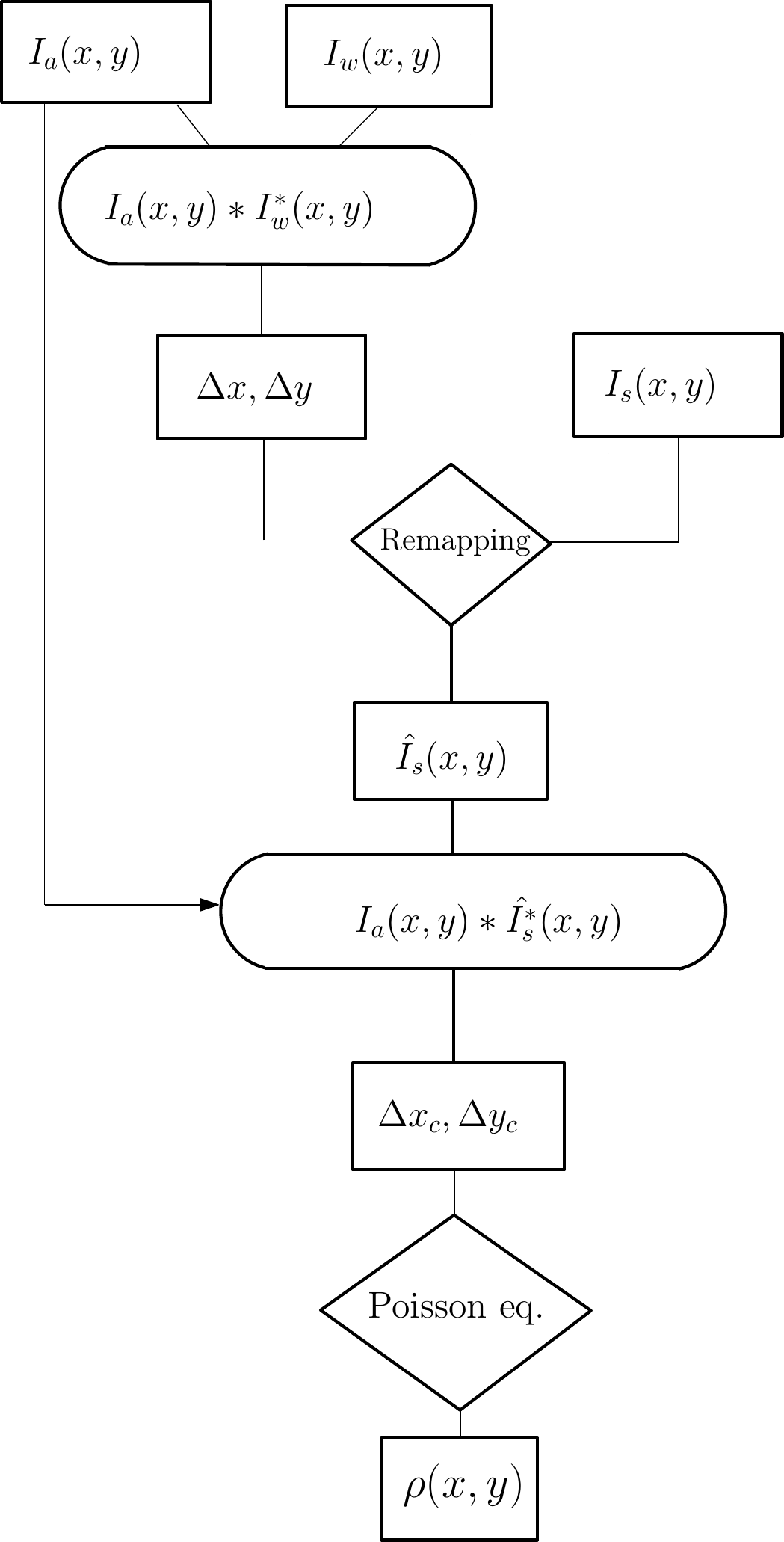}
\caption{Flow chart of the procedure.\label{fig_flowchart}}
\end{figure}

The order of the steps are shown in a block diagram in Fig.\ref{fig_flowchart}:
\begin{itemize}
\item We capture three images of the background pattern, through air $I_a(x,y)$, water $I_w$ and a saline stratified solution, $I_s$. ($I$ stands for image). 
\item The first calibration is the displacement field $\Delta x, \Delta y$ obtained correlating the air and water images, $I_a \ast I^*_w$, where $\ast$ is a convolution operator and subscript $^*$ is the conjugate (reversed) image. 
\item The background pattern image obtained through the saline stratified solution is first remapped using the displacement field whose origins are in the optical system and aberrations due to the multi-media (air-glass-water-glass-air) imaging
\item The corrected image $\hat{I}_s$ is correlated with the original reference image taken in air $I_a$, and the result $\Delta x_c, \Delta y_c$ is used to construct and solve the Poisson equation. 
\item The result of the Poisson equation solution is the desired density field $\rho(x,y)$ (applying the  $n \to \rho$ conversion.) 
 \end{itemize} 

\begin{figure}[ht]
\centering{}\includegraphics[width=.7\textwidth]{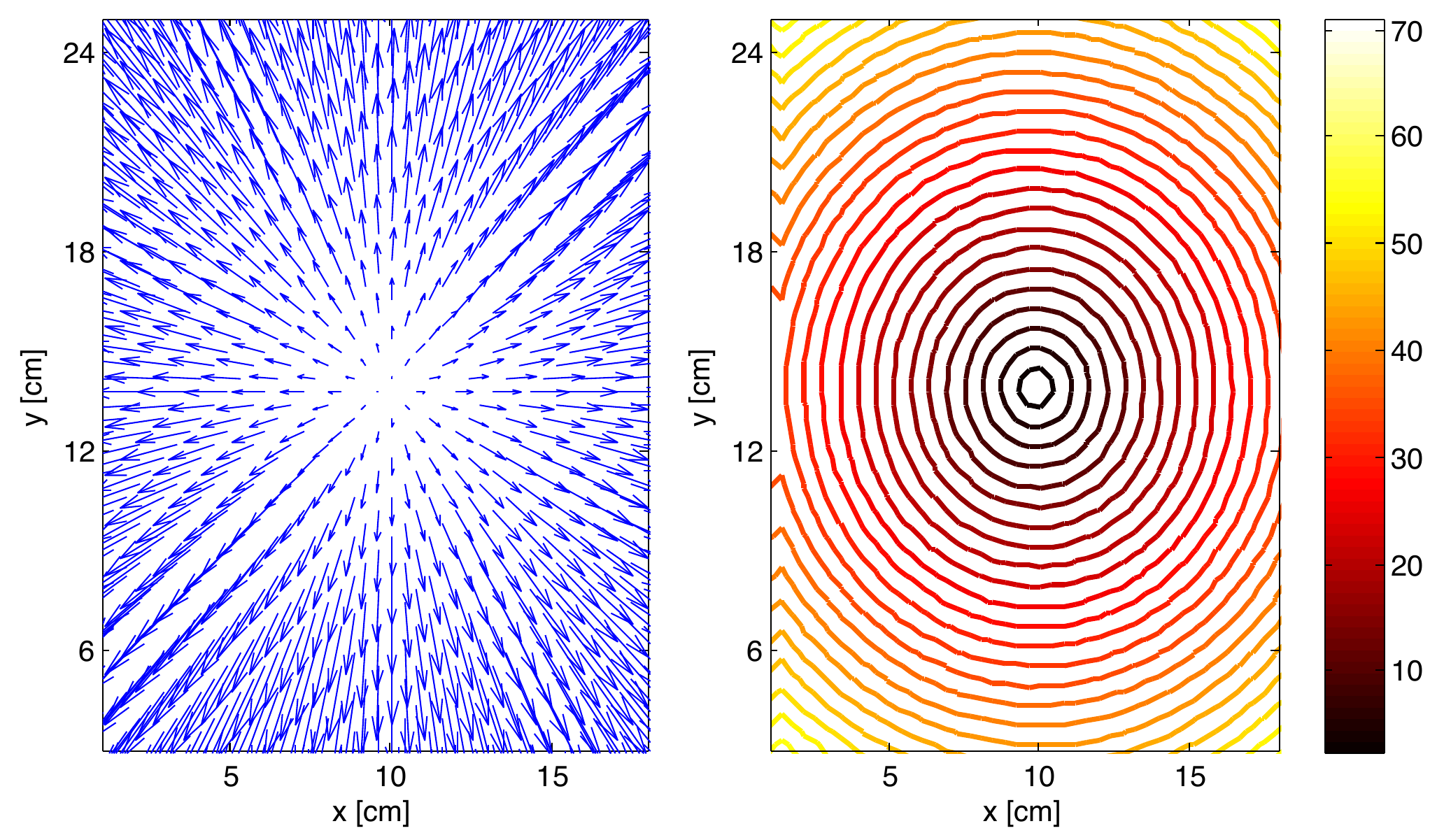}
\protect\caption{(Left) Crop of the displacement vector field assumed as a calibration and (Right) contour of its respective magnitude values. \label{fig_table_cor}}
\end{figure}

The result of single steps are shown in Fig.~\ref{fig_table_cor}. The left plot shows the displacement field $\Delta x, \Delta y$ as arrows. 
The plot emphasizes a typical distortion created by the multi-media imaging of the background pattern. The  displacement field is quantified by cross-correlating the air and water images, $I_a$ and $I_w$. The two images were analyzed using a standard FFT-based correlation method with an interrogation area of $ 16 \times 16$ pixels$^2$.  The magnitude of the displacement field $\sqrt{\Delta x^2 + \Delta y^2}$ is shown as contours in the right plot of Fig.~\ref{fig_table_cor}. The maximum displacement is tens of pixels and it is larger at the edges of the image. Note that we present the whole field measurement method and the image corresponds to the field of approximately $20 \times 25$ cm$^2$. Seventy pixels distortion is not visible in the four megapixel image. However, we understand that since the Poisson equation utilizes the integration from the edges of boundary conditions, the error propagates to the place where the result is important.

In order to correct the image distortion, the image remapping code was developed. First, a calibration field is obtained by correlating the reference image and the image of the background pattern through the full  tank of water, $I_w$. In the following stage, this calibration field is used to correct the distortion generated by the saline water image.  The calibration vector field is re-sampled on a dense rectangular mesh grid using linear interpolation. The resulting new field is then applied to each pixel of the saline solution image using the standard image remapping method: $\hat{I} = I(T_x(x,y),T_y(x,y))$ ($T$ denotes the transform map).   It  remaps each pixel of the distorted image, inverting the displacement field of the calibration image.  

\begin{figure}[ht]
\centering
\includegraphics[width=0.8\textwidth]{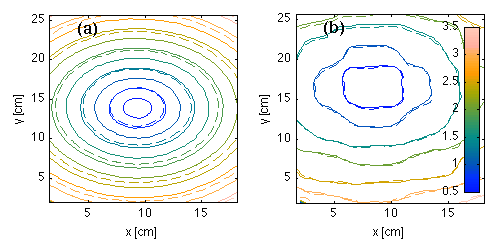}
\caption{(a) Contour plot of the displacement magnitude for the air-water (solid line) and air-saline solution (dashed lines). (b) Comparison of the displacement magnitude of water-saline solution (solid) and corrected, remapped version (dashed). \label{fig_comp_panel}}
\end{figure}

The effect of remapping is not significant and therefore not easily visualized. We demonstrated the contour maps of displacement fields with and without correction for the: a) air-water case; and b) multi-layer stratified solutions in Fig.~\ref{fig_comp_panel}. The original, not remapped results, are shown by solid contours and the corrected ones by the dashed lines. The contour maps are very similar. Nevertheless, as we explained above, the error accumulates during the solution of the Poisson equation. As the next section shows, the results are very significant for the reconstructed density field. 

\section{Experimental results}\label{sec:setup}

The experiments are performed in a glass tank with a $20 \times 20$
cm$^{2}$ cross-section and a height of 30 cm. 
The random dot pattern is created using a Matlab script (makebospattern.m courtesy of Frederic Moisy, \url{http://www.fast.u-psud.fr/pivmat}). There are 200,000 black dots distributed randomly over an A4 size transparent sheet.  The transparent sheet with a background pattern was attached at the back-side of the tank. It is illuminated with a white LED light, equipped with a plastic light diffusing sheet.  The light distribution is approximately uniform. The non-uniformity of the light due to the lack of parabolic mirrors in the digital BOS application is corrected by the proposed method.  

The imaging system uses a four megapixel CCD camera (Optronis CL4000CXP) with a 10-bit sensor of $2304  \times 1720$ pixels that yields a magnification of 56.2 $\mu$m/pixel.  All the BOS image pairs were processed similarly to the PIV images (in the present case using FFT-based cross-correlation of $16 \times 16$ pixels).

We present here two important tests, namely Test I and II.  In Test I, we use the reference image in air $I_a$ only, and implement the method to obtain the position of the air-water interface and reconstruct the density of the two fluids. The air-water interface was shifted between different runs, removing the water from the tank through a valve.    

In Test II, we implement the method using a reference image in air, a reference image in water $I_w$, and attempt to reconstruct the density field of the stratified solution of water/Epsom salt ($MgSO_{4}$). In order to emphasize the accuracy of the method, we establish four layers of distinct density difference, $\rho = 1.00, 1.12, 1.20, 1.27 $ g/ml (using a calibrated pycnometer). Each layer, in addition, is naturally stratified.  The results of the two tests are presented below.

\subsection{Test I results}

\begin{figure}[ht]
\centering
\includegraphics[width=.9\textwidth]{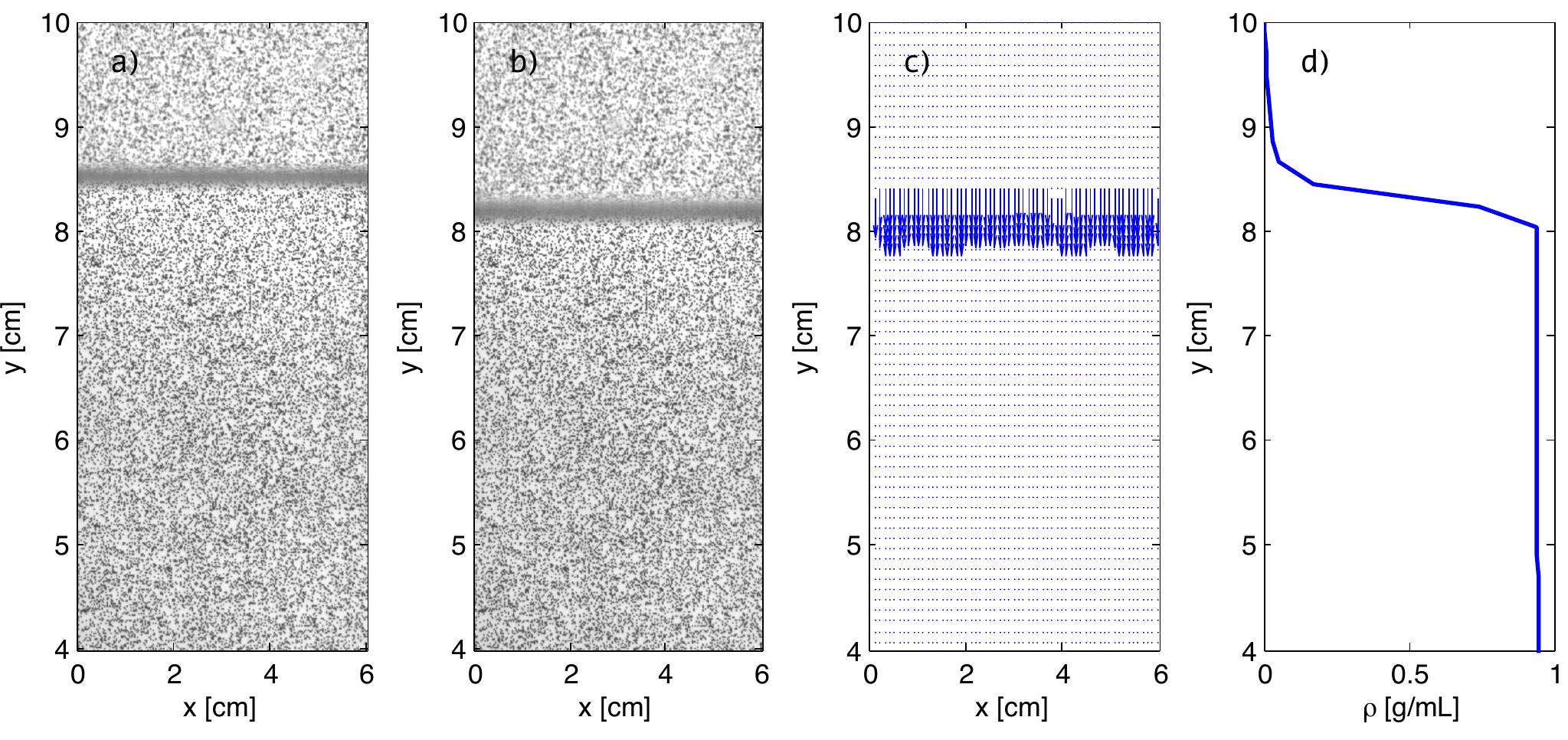}
\caption{Results of Test I. (a)-(b) images of the background pattern through two layers of air and water. The dark line is the interface at two different positions. (c) The displacement field of the two images. (d) The density profile reconstructed via the Poisson equation. \label{fig_shift}}
\end{figure}

Two example images obtained in Test I (air-water interface) are shown in Fig.~\ref{fig_shift}.  The top panel shows the images of the background pattern. A dark line is the interface between air (top part) and water (bottom part). We tested several positions of the air-water interface (not shown here for the sake of brevity). Fig.~\ref{fig_shift}c  shows the result of the first step in the BOS method, the correlation of these images with a reference image. The largest displacements appear at the interface. It is important to note that the results depend on a relative height of the air-water interface with respect to the imaging axis of the camera (i.e. the angle between the ray and the imaging axis in  Fig.~\ref{fig_system}). 
The result of the algorithm is the density field $\rho(x,y)$ from two layers, $\rho_a$ and $\rho_w$ for air and water, respectively. The density field is homogeneous in the horizontal direction and the result is shown as a spatial average along $x$ in Fig.~\ref{fig_shift}d.  The position of the interface is shown by a jump from the density of air $\rho_a$ to $\rho_w$. 

\subsection{Test II results}

\begin{figure}
\centering\includegraphics[width=.53\textwidth]{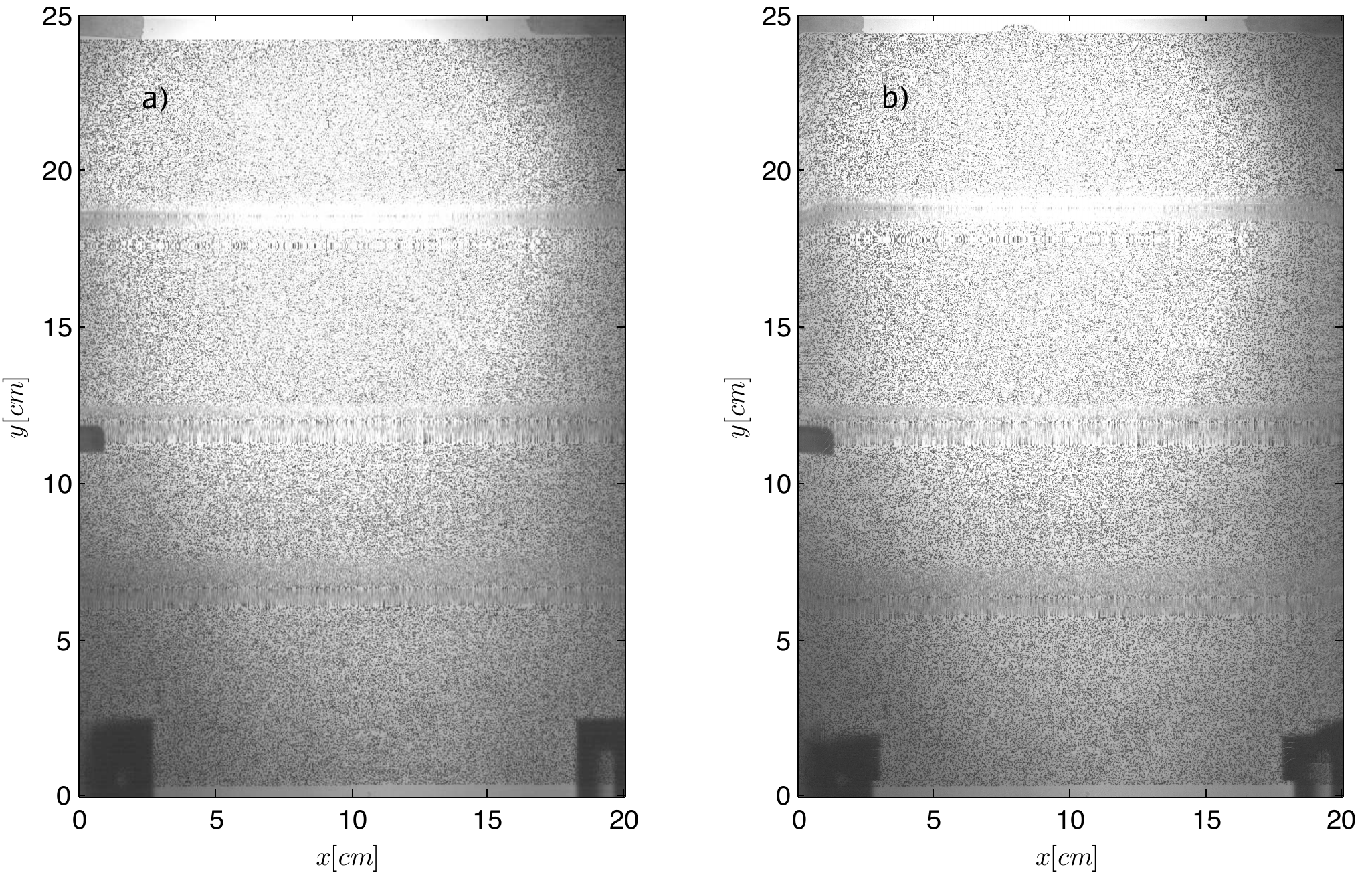}
\includegraphics[width=.45\textwidth]{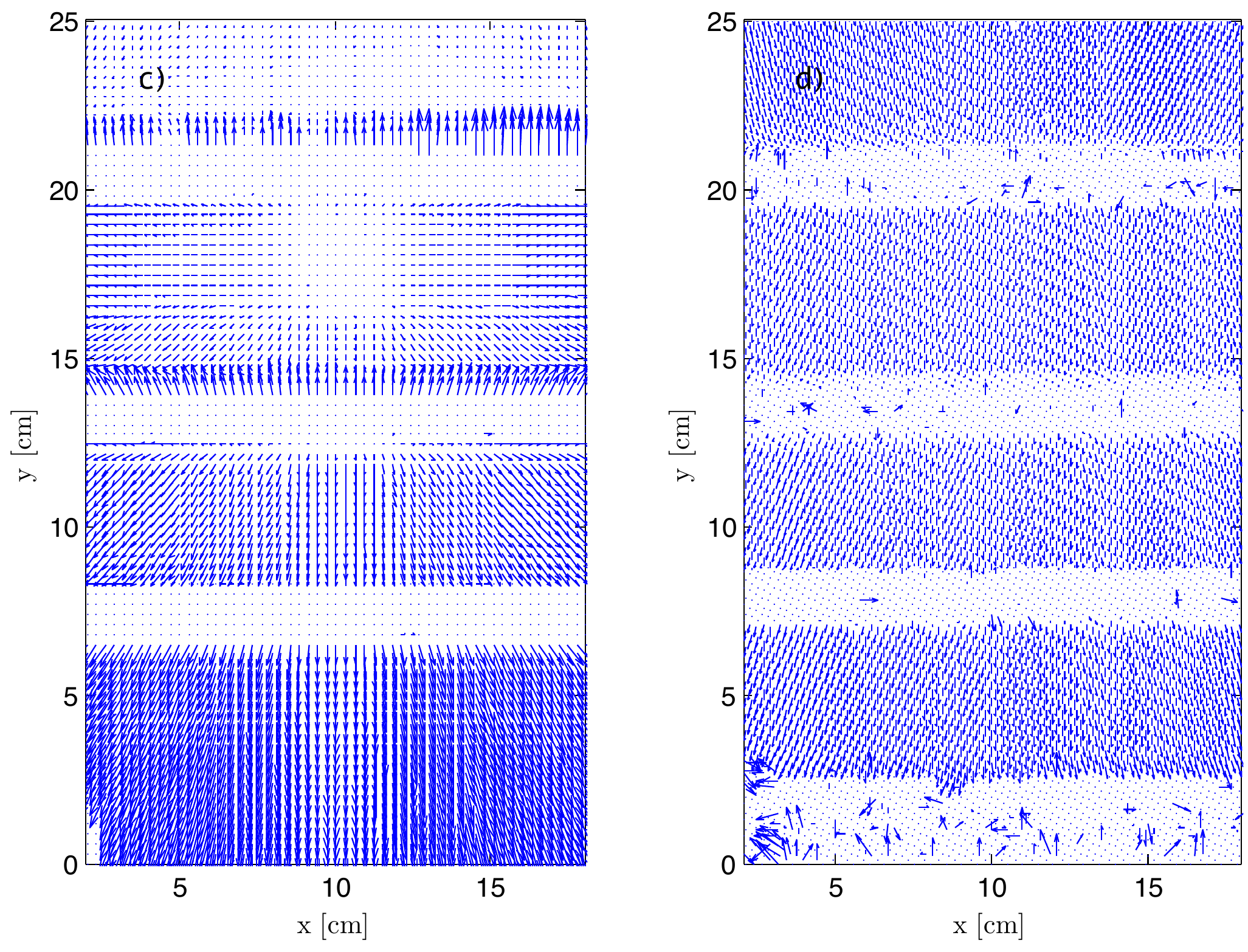}
\caption{(a) Original image of the BOS pattern through the four layers of increasing density of saline water solutions. (b) Remapped image. (c) Correlation of the water reference image with the 4 layers image. (d) Correlation result of the remapped image.\label{fig_original_4layers} }
\end{figure}

In Test II we measure the density field of the four layers of stratified saline solution. The original image is  shown in  Fig.~\ref{fig_original_4layers}a-b together with the remapped image. The effect of the remapping algorithm is not clearly seen. However, after the BOS calculations we observe a striking difference. Utilizing the naive BOS method to the air-saline solution creates an artifact.  For the sake of completeness, we first plot the result of the correlation in  Fig.~\ref{fig_original_4layers}c, compared with the result of the remapped (corrected) image pair in  Fig.~\ref{fig_original_4layers}d.  

\begin{figure}[ht]
\centering
\includegraphics[width=.95\textwidth]{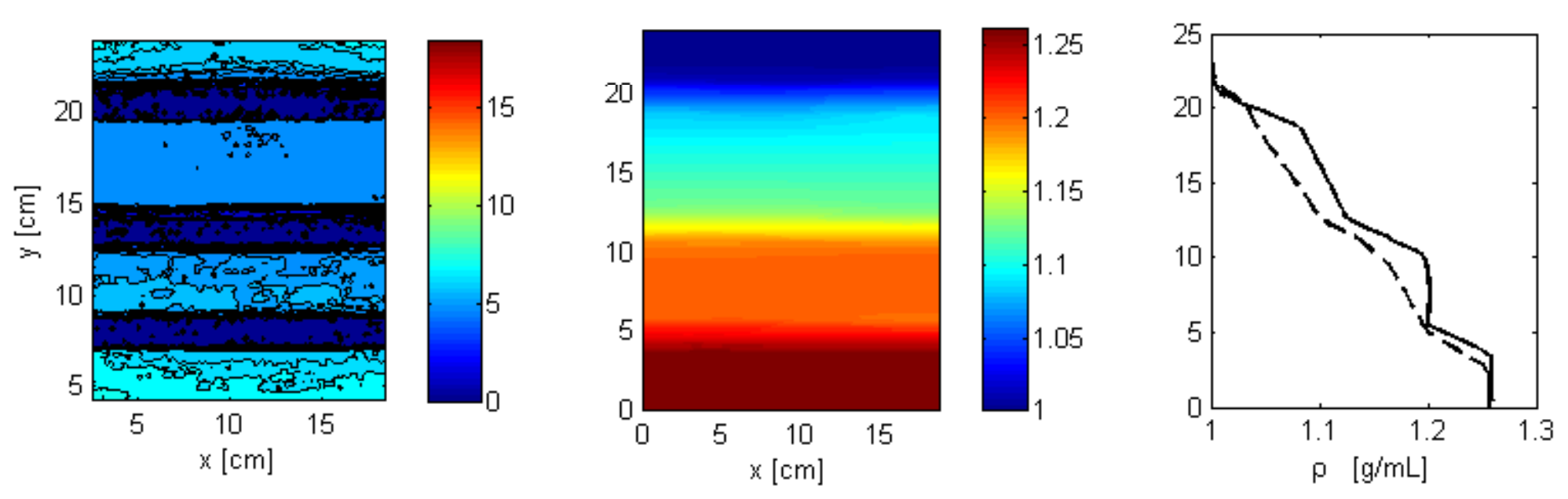}
\caption{Results of the corrected image pair analysis (left) the magnitude of the displacement field, $(dx^2+dz^2)^{1/2}$, (center) the Poisson equation solution. Color maps correspond to the displacement field in pixels and the density in [g/ml], respectively (right). The final result of the BOS method, the density profile $\rho(z)$ for the four layer saline solution. Solid line - corrected solution using the new method, dashed --the Poisson solution of the water-saline solution image pair displacement field without correction. \label{fig_4layers_profile}}
\end{figure}

The final and the major result of the BOS method presented in this work is shown in Fig.~\ref{fig_4layers_profile}. The plot shows the magnitude of the displacement field that is used in the Poisson equation. Next we demonstrate the solution of the Poisson equation, converted to the density units. The right panel demonstrates the spatially averaged profile of density. The solid line corresponds to the result shown in Fig.~\ref{fig_4layers_profile} that uses the remapped, corrected pair of images. The dashed line represents the result of the water-saline solution pair analysis without correction.  Although at the boundaries the values are correct by definition, obviously the non-corrected image pair leads to a completely wrong density profile. We verify that the reconstruction method provides the correct positions of the layers of different density. The results in our case show that not only the density layers are accurately reconstructed, but also the actual values are within a 5\% error. 

\section{Summary and conclusions}\label{sec:conclusions}

Literature reviews show an increase of applications of optical 
measurement methods of density fields based on the index of refraction of the fluid. This is partially due to the progress in imaging technology. It promotes the use of synthetic, digital optical methods, such as synthetic Schlieren. In most cases, the methods applied to gas flows or non-stratified liquid flows. Generally the result of the density gradient field is sufficient. Obtaining the density field could be difficult due to the phenomenon disclosed in this work -- the stratification affects the accuracy of reconstruction of the density field. For instance, in the setup using water and saline solutions, the changes of the index of refraction are modified due to stratification. The consequence of the density differences in the stratified flow is the distortion of images. For some
optical measurement techniques, such as particle image velocimetry (PIV), the variation of refractive index
in the bulk of fluid is just a source of additional error. Applying the background oriented Schlieren (BOS) method to the PIV result of distorted images results in a completely wrong density field. 

In this paper, the method to reconstruct the density field in  stratified
liquid flows using BOS is proposed. 
To the best of our knowledge, the reconstruction of the correct density field in multi-layer stratification is performed  for the first time. The method works only for stagnant fluid in which the density field is two dimensional, without density gradients along the imaging axis.

By analyzing differences in images of background pattern in air and water, we identified the cause of image distortion. There is an apparent displacement of the dots due to refractive index variations. These have a non negligible
effect on the magnitude and orientation of the displacement vector field. The error is especially high in the corners of the images, partially due to imaging optics. The distortion then propagates into the final result through the solution of the Poisson equation. 

The algorithm developed in this work corrects the images using deconvolution methods. The correction reduces the measurement
errors and allows for quantitative density measurements in stratified fluids. In addition, this correction is useful for the non-perfectly parallel light sources and digital imaging optics. 

Our method improves the applicability of BOS. The corrected images enable producing quantitative density field data, comparable with direct, intrusive, density measurements. Hopefully, this method will increase the application of BOS to the stratified flows. There is an additional study required to verify the accuracy of the method of gaseous and liquid flows of strongly changing index of refraction. 



\begin{thebibliography}{10}
\providecommand{\url}[1]{{#1}}
\providecommand{\urlprefix}{URL }
\expandafter\ifx\csname urlstyle\endcsname\relax
  \providecommand{\doi}[1]{DOI~\discretionary{}{}{}#1}\else
  \providecommand{\doi}{DOI~\discretionary{}{}{}\begingroup
  \urlstyle{rm}\Url}\fi

\bibitem{Berger}
Berger, K., Ihrke, I., Atcheson, B., Heidrich, W., Magnor, M.A.: Tomographic 4d
  reconstruction of gas flows in the presence of occluders.
\newblock VMV \textbf{DNB}, 29--36 (2000)

\bibitem{Dalziel2000}
Dalziel, S., Hughes, G., R., S.B.: {Whole-field density measurements by
  synthetic schlieren}.
\newblock Exp. Fluids \textbf{28}, 322--335 (2000)

\bibitem{Elsinga}
Elsinga, G., Van~Oudheusden, B., Scarano, F., Watt, D.: {Assessment and
  application of quantitative schlieren methods: calibrated color schlieren and
  background oriented schlieren}.
\newblock Exp. Fluids \textbf{36}, 309--325 (2004)

\bibitem{Goldstein}
Goldstein, R.: Optical systems for flow measurement:shadowgraph, schlieren, and
  interferometric techniques.
\newblock Hemisphere Publishing Corp. (1983)

\bibitem{Raffel2015}
Raffel, M.: {Background-oriented schlieren (BOS) techniques}.
\newblock Exp. Fluids pp. 56--60 (2015)

\bibitem{Richard1998}
Richard, H., Raffel, M.: Background oriented schlieren demonstrations.
\newblock Final report for the European Research Office Edison House, Old
  Marylebone Road London NW15TH, United Kingdom pp. 223--231 (1998)

\bibitem{Settles}
Settles, G.: Schlieren and shadowgraph techniques : visualizing phenomena in
  transparent media.
\newblock Springer, Berlin (2001)

\bibitem{Sutherland:1999}
Sutherland, B., Daziel, S., Hughes, G., Linden, P.: Visualization and
  measurement of internal waves by synthetic schlieren. {Part 1}. vertically
  oscillating cylinder.
\newblock Dynamics of Atmospheres and Oceans \textbf{390}, 93--126 (1999)

\bibitem{tropea_yarin_foss:2007}
Tropea, C., Yarin, A.L., Foss, J.F.: {Springer Handbook of Experimental Fluid
  Mechanics}.
\newblock Springer, Berlin, Heidelberg (2007)

\bibitem{Venkatakrishnan2004}
Venkatakrishnan, L., Meier, G.E.A.: {Density measurements using the Background
  Oriented Schlieren technique}.
\newblock Exp. Fluids \textbf{37}, 237--247 (2004)

\bibitem{Vinnichenko}
Vinnichenko, N., Uvarov, A., Plaksina, Y.: {Accuracy of background oriented
  schlieren for different background patterns and means of refraction index
  reconstruction}.
\newblock In: 15th international symposium flow visualization, Minsk (2012)

\end{thebibliography}

\end{document}